\documentclass{JINST}
\RequirePackage{xspace}
\RequirePackage{bm}
\def\epem {\ensuremath{e^+e^-}\xspace}
\def\lplm {\ensuremath{l^+l^-}\xspace}
\def\gg {\ensuremath{\gamma \gamma}\xspace}
\def\ge {\ensuremath{\gamma e}\xspace}
\def\invpb {\ensuremath{\mathrm{pb}^{-1}}\xspace}

\newcommand{\kev}{\ensuremath{\mathrm{\,ke\kern -0.1em V}}\xspace}

\def\mum  {\ensuremath{{\,\mu\rm m}}\xspace}

\newcommand{\gev}{\ensuremath{\mathrm{\,Ge\kern -0.1em V}}\xspace}
\newcommand{\gevc}{\ensuremath{{\mathrm{\,Ge\kern -0.1em V\!/}c}}\xspace}
\newcommand{\gevcc}{\ensuremath{{\mathrm{\,Ge\kern -0.1em V\!/}c^2}}\xspace}
\newcommand{\kevcc}{\ensuremath{{\mathrm{\,ke\kern -0.1em V\!/}c^2}}\xspace}

\newcommand{\GG}{\mbox{$\gamma\gamma$}}
\newcommand{\GE}{\mbox{$\gamma e$}}

\newcommand{\be}{\begin{equation}}
\newcommand{\ee}{\end{equation}}
\newcommand{\bc}{\begin{center}}
\newcommand{\ec}{\end{center}}
\newcommand{\bi}{\begin{itemize}}
\newcommand{\ei}{\end{itemize}}
\newcommand{\ben}{\begin{enumerate}}
\newcommand{\een}{\end{enumerate}}

\title{Energy calibration at high-energy photon colliders }

\author{V.~I.~Telnov  \\
Budker Institute of Nuclear Physics,\\
Novosibirsk State University, \\ 630090, Novosibirsk, Russia\\
E-mail: \email{telnov@inp.nsk.su}}

\abstract{Calibration of the absolute energy scale at high-energy photon (\gg, \ge) colliders is discussed. The luminosity spectrum at photon colliders is broad and has a rather sharp high-energy edge, which can be used, for example, to measure the mass of the Higgs boson in the process $\gg \to H$ or masses of charged scalars by observing the cross-section threshold.  In addition to the precise knowledge of the edge energy of the luminosity spectrum, it is even more important to have a way to calibrate the absolute energy scale of the detector. At first sight, Compton scattering itself provides a unique way to determine the beam energies and produce particles of known energies that could be used for detector calibration.  The energy scale is given by the electron mass $m_e$  and laser photon energy $\omega_0$. However, this does not work at realistic photon colliders due to large nonlinear effects in Compton scattering at the conversion region ($\xi^2 \sim 0.3$). It is argued that the process $\ge \to eZ_0$ provides the best way to calibrate the energy scale of the detector, where the energy scale is given by $m_Z$.  }

\keywords{Instrumentation for particle accelerators and storage rings - high energy (linear accelerators, synchrotrons); Performance of High Energy Physics Detectors; Pattern recognition, cluster finding, calibration and fitting methods; Detector alignment and calibration methods (lasers, sources, particle-beams)}

\begin{document}

\section{Introduction}\label{s1}
From \epem storage-ring experiments, we know that precise knowledge of the beam energy is very useful as it enables the determination of particle masses with fantastic precision, practically independent of the detector resolution and its systematic errors. The method of resonant beam depolarization at storage rings has enabled the measurement of $M_Z$ at LEP with a relative accuracy of $2.3\cdot10^{-5}$, and $M_{J/\Psi}$\cite{PDG} was measured at Budker INP with a relative precision of $4 \cdot 10^{-6}$ ($\sigma_M = 12 \kevcc $!)\cite{KEDR}.

At linear \epem colliders, there is a desire to determine the absolute beam energy using special magnetic spectrometers upstream and downstream from the interaction point (IP) with an accuracy $\sigma_E/E \sim 10^{-3}$ and $10^{-4}$, respectively~\cite{ILC-spec}. The luminosity-weighted center-of-mass energy can be found using the radiative-return $Z$ production: $\epem \to Z\gamma \to \mu^+ \mu^-\gamma $, where $\gamma$ travels at a small angle relative to the beam direction and is not detected in most cases. Since the $M_Z$ is well-known, the c.m.s. energy can be reconstructed by measuring only the angles of muons. This method was successfully used at LEP-2 and can be used at linear \epem colliders as well. The expected relative accuracy of this method is $10^{-4}$ for $2E_0=350$ GeV and an integrated luminosity of 100 fb$^{-1}$~\cite{Monig}.   At linear \epem colliders, the beam energy spread is about 0.15\%. During the beam collision, a large fraction of beam particles emit beamstrahlung and ISR photons; nevertheless, a narrow spike in the luminosity spectrum remains. It can be used in measurement of particle masses and fine structures in cross sections such as the $t$-quark threshold, SUSY thresholds, $Z^{\prime}$, etc. Using energy scanning with a narrow luminosity spectrum, one can measure particle masses much better than they could be measured by the detector.

\begin{figure}[hbt]
\centering
\vspace*{-0.6 cm}
\hspace{-0.5cm}   \includegraphics[width=10cm,angle=0]{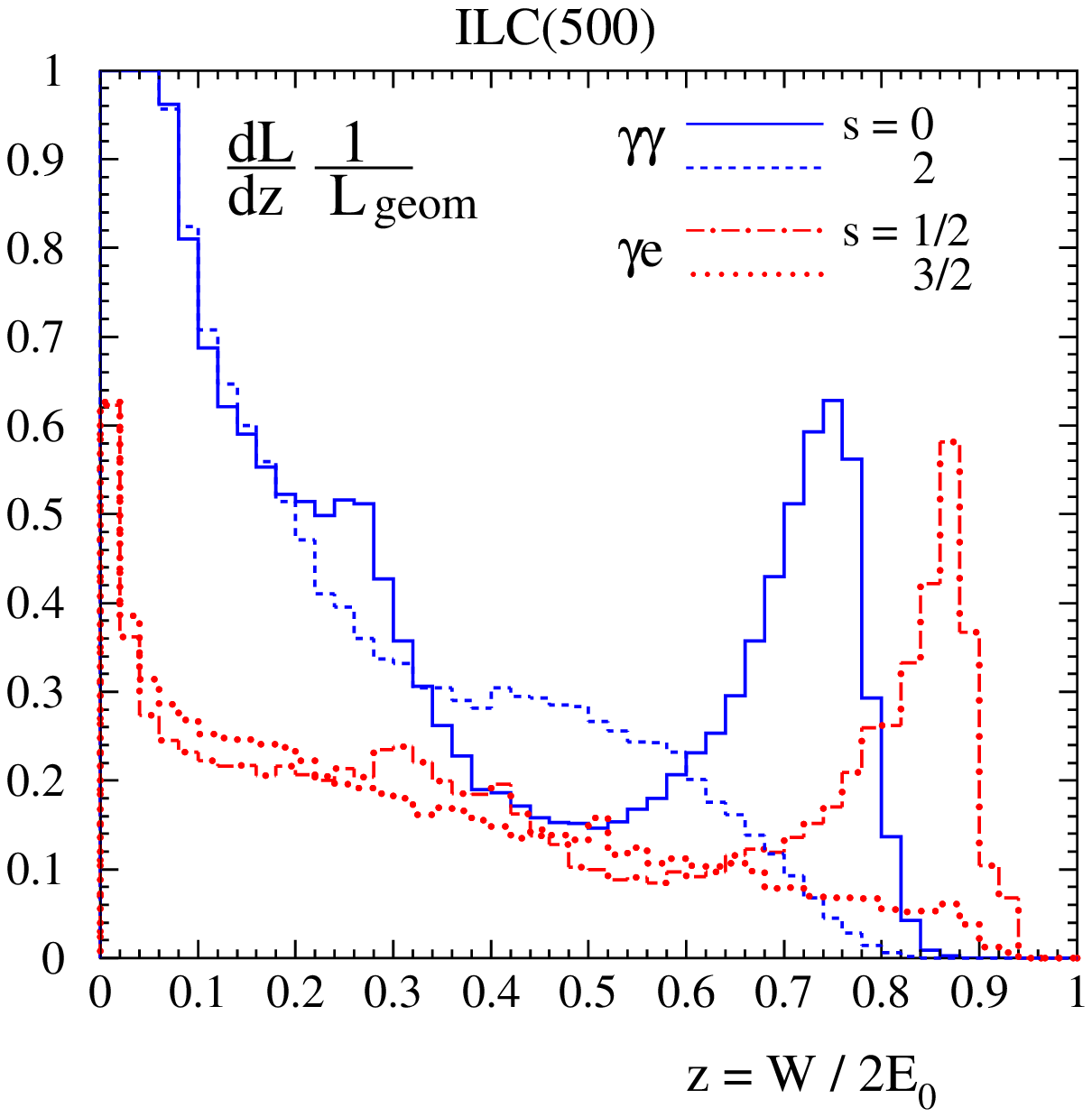}
\vspace{-0.8cm}
\caption{ \GG, \GE\ luminosity spectra for ILC at $2E_0=500$ GeV.}
 \vspace{-0.0cm}
\label{lumspectra}
\end{figure}

Luminosity spectra at photon colliders (PC), see Fig.~\ref{lumspectra}~\cite{TESLA-TDR,Telnov2006}, are quite broad with rather sharp edges, which can be useful for measurement of particle masses. Since photons have wide
spectra and various polarizations, in general case, one has to measure 16 two-dimensional luminosity distributions $d{\,^2}L_{ij}/d\omega_1
d\omega_2$,  $dL_{ij} = dL_{\gg} \langle \xi_i \tilde{\xi_j} \rangle$,
where $\xi_i$ are the Stokes parameters of the photons and the tilde denotes the second colliding beam. Among the 16 cross sections $\sigma_{ij}$, three are the most important: those that do not vanish after averaging over the spin states of final particles and the azimuthal angles. These luminosity spectra can be found experimentally using polarization-sensitive QED processes~\cite{Pak}. In order to measure the spectra with sharp edges, the tracking system must have good momentum resolution. The expected resolution of tracking systems at ILC detectors $\sigma_p/p \sim 3\cdot 10^{-5}p\,[\gevc]$ for at $p>100 \gevc$~\cite{ILC}, which is sufficient.  In principle, using energy scanning by the sharp edge of the luminosity spectra, one can measure the Higgs mass with an accuracy better than $10^{-3}$~\cite{Telnov98,Ohgaki}. Also, photon colliders would have an advantage in the measurement of charged scalar masses using the energy scanning near the threshold because $\sigma(\gg \to S^+S^-) \propto \beta$, while $\sigma(\epem \to S^+S^-) \propto \beta^3$. However, in a majority of physics studies the reconstruction of events will be based on information from the detector tracking system and calorimeters, which should be properly calibrated.

\section{Nonlinear effects in Compton scattering}\label{sec:yyy}
The maximum photon energy after Compton scattering of a laser photon with the energy $\omega_0$ on a high-energy electron with the energy $E_0$~\cite{GKST83}
\be
\omega_m=E_0\frac{x}{x+1}, \;\;\;{\rm where}\;\;\;x=\frac{4E_0\omega_0}{m_e^2c^4},
\ee
This expression is valid only for low laser intensities (linear Compton scattering). For $2E_0=500$ GeV and $\lambda=1$ \mum, the parameter $x\sim 4.5$. In order to get a conversion coefficient close to 100\%, the density of laser photons should be so high that several laser photon can interact simultaneously, leading to nonlinear effects in Compton scattering\cite{Berestetskii,TESLA-TDR}. These effects are characterized by the parameter $\xi^2=e^2\bar{B}^2\hbar^2/m_e^2\omega_0^2c^2=2n_{\gamma}r_e^2\lambda/\alpha$, where $\bar{B}$ is the r.m.s.\ strength of the electric (magnetic)
field in the laser wave, $n_{\gamma}$ is the density of laser photons. In a strong laser field, the electron acquires the effective mass $m_e^2 \to m_e^2(1+\xi^2)$, and therefore $x\to x/(1+\xi^2)$.  This leads to the decrease of the maximum photon energy of scattered photons
\be
\omega_m=E_0\frac{x}{x+1+\xi^2},\;\;\;\;\;\frac{\Delta \omega_m}{\omega_m}\approx -\frac{\xi^2}{x+1} \;\;{\rm for} \;\;\; \xi^2 \ll x.
\ee
For $\xi^2=0.3$ and $x=4.5$, the energy decreases by 5\%. Thus, this criteria determines the acceptable $\xi^2$ values. Obtaining smaller $\xi^2$ given a fixed conversion coefficient requires a larger laser flash energy~\cite{TESLA-TDR}, which is technically problematic.

Additionally, the density of photons at the laser focus varies, which results in
 a spread $\sigma_{\xi^2} \sim 0.4\langle\xi^2\rangle \sim 0.28\xi^2(0)$ (obtained by simulation).  If the average shift is 4\%, then the additional r.m.s.\ energy spread is 1.5\%.  So, the high-energy edge of the $\gamma$ spectrum is not very sharp (slope $\sim$ 3--4\%), and the maximum energy is unstable due to the possible variations of the laser focus geometry (displacement, change of the spot size). The edge photon energy could be measured in \ge collisions where, in the ideal case (low laser intensities), the luminosity spectrum has a very sharp edge; however, due to nonlinear effects in Compton scattering, it has a $\sim$1.5\%
energy spread plus some additional spread due to possible variations of laser intensity at the laser focus.

We have shown that 

\begin{enumerate}
  \item At photon colliders, the main uncertainty in the energy of colliding photons is caused by the uncontrolled variation of laser intensity in the conversion region. The ratio of the maximum photon energy to the electron beam energy is not a constant due to nonlinear effects in Compton scattering; therefore, it can not be used to determine the beam energy.
  \item The characteristic spread (width) of the high-energy edge of luminosity spectra is about 3--4\%, which is greater than the detector resolution ($\sim 0.3$ \% at $E=100$ GeV). The relative luminosity spectrum can be measured using QED processes ($\gg \to \lplm, \gg \to \lplm \gamma, \ge \to \ge, \ge \to e\epem, \ge \to eZ$, etc.)~\cite{Pak}.
      \item The knowledge of the electron beam energy does not help too much; calibration of the absolute energy scale of the detector is needed.
  \end{enumerate}

  \section{Calibration of the detector}
  \subsection{$\ge \to \ge$}
  In order to measure energy, one needs some value with a dimension of mass. At first sight, one can use $\ge$ collisions (the energy scale is given by the electron mass $m_e$).  The scattering angles in collisions of electrons with energy $E_0 $ and photons with edge energy $\omega_m$ allow one to determine $x=4E_0\omega_0/ m^2c^4$, and thus to find $E_0$. For linear (low-intensity) Compton scattering, the ratio of the maximum photon energy after the $e\to \gamma$ conversion and the electron beam energy can be found by measuring the angles of scattered photon and electron (angles with respect to the initial electron direction):
  \be
 \frac{\omega_m}{E_0}=\frac{x}{x+1}=\frac{\sin\theta_1+\sin\theta_2-\sin(\theta_1+\theta_2)}{\sin\theta_1+\sin\theta_2+\sin(\theta_1+\theta_2)}.
  \ee
By measuring the edge in distribution of this parameter, one can find the value of $x$, and then the beam energy.  However, as we saw above, due to nonlinear effects in Compton scattering this measurement gives not $x$, but $x/(1+\xi^2)$, and due to large uncertainty in $\xi^2$ the accuracy of the beam energy determination will be very poor:
  \be
  \frac{\sigma_{E_0}}{E_0}\sim \frac{\sigma_{\xi^2}}{1+\xi^2}\sim {\mathcal{O}(1 \% )}.
  \ee

\subsection{$\ge \to eZ$}
\begin{figure}[hbt]
\centering
\vspace*{-0.0 cm}
\hspace{-0.5cm}   \includegraphics[width=13.cm,angle=0]{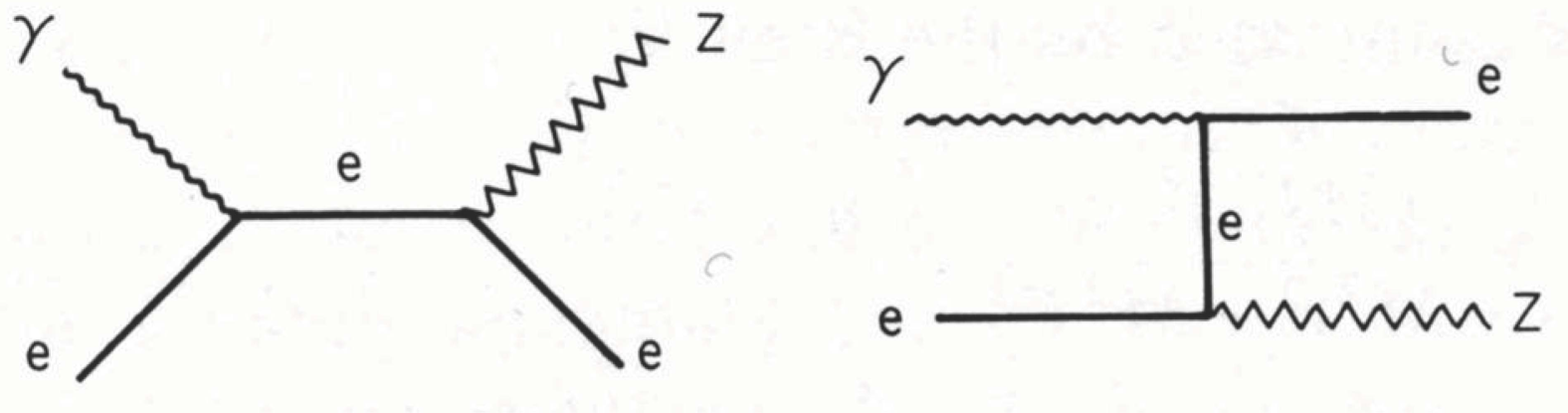}
\vspace{-0.cm}
\caption{ The process $\ge \to eZ$.}
 \vspace{-0.0cm}
\label{eZ}
\end{figure}

   Here, the energy scale is given by $M_Z$. The diagrams for this process are shown in Fig.~\ref{eZ}. The second diagram dominates; the $Z$ boson travels predominantly in the direction of the initial electron (the process can be viewed as $\epem \to Z$ annihilation after $\gamma \to \epem$ virtual decay). The dominant term in the angular distribution~\cite{Renard}
   \be
   \frac{d\sigma}{d \cos\theta_Z} \propto \frac{1}{1-\cos\theta_Z+ 2m^2_e/W^2}.
   \ee
    In most cases, only $Z$ decay products are detected (for example, $\mu^+\mu^-$), and the final-state electron escapes at a small angle. In this case, the energy of the initial electron can be express via the $Z$ mass and the angles $\theta_{1,2}$ between the muons and direction of the initial electron:
\be
E_0\approx 0.5M_Z\sqrt{\frac{\sin\theta_1+\sin\theta_2+\sin(\theta_1+\theta_2)}{\sin\theta_1+\sin\theta_2-\sin(\theta_1+\theta_2)}}.
\ee
  This expression is similar to that for the process $\epem \to Z\gamma$, which is used for energy determination in \epem collisions~\cite{Monig}.

\begin{figure}[hbt]
\centering
\vspace*{-0.0 cm}
\hspace{-0.cm}   \includegraphics[width=13.cm,angle=0]{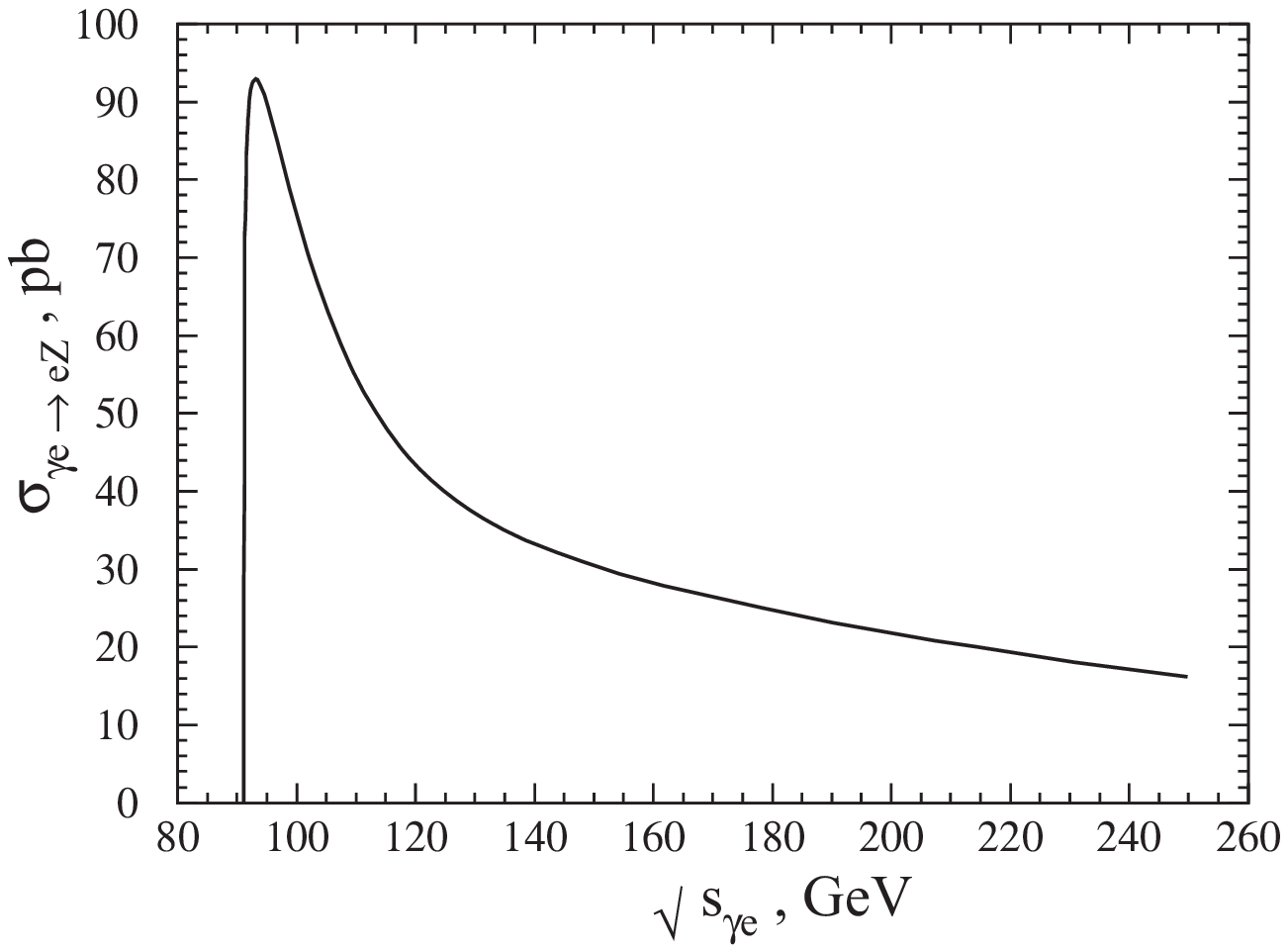}
\vspace{-0.cm}
\caption{ The cross section $\sigma_{\ge \to eZ}$.}
 \vspace{-0.0cm}
\label{eZ-cross}
\end{figure}

  The cross section for unpolarized beams $\ge \to eZ$ is given by~\cite{Renard,Ginzburg}:
  \be
  \sigma_{\ge\to Ze}=\frac{\tilde{\sigma}}{x}\left[\left(1-\frac{2}{x} +\frac{2}{x^2}\right) L+\frac{1}{2}\left(1-\frac{1}{x}\right)\left(1+\frac{7}{x}\right)\right], \;\;\;x=\frac{s_{\ge}}{M_z^2},
  \ee
  $$\tilde{\sigma}=\frac{\pi\alpha^2}{2M_Z^2\sin^22\theta_W}\left[1+(4\sin^2\theta_W-1)^2\right]=5.9 \mathrm{~pb} $$
  $$L=\ln\frac{(s_{\ge}-M_Z^2)^2}{m^2_e s_{\ge}}\approx 24 +\ln\frac{(x-1)^2}{x}$$
  The total cross section is shown in Fig.~\ref{eZ-cross}.

Although this method allows the determination of the beam energy, it does not make much sense because 
 \bi
 \item due to Compton scattering and beamstrahlung, only a rather small number of electrons have the initial beam energy $E_0$;
 \item the knowledge of the beam energy is not particularly useful because, due to nonlinear effects, the maximum photon energy after Compton scattering is not perfectly related to the electron energy.
 \ei
 What is really needed is a source of $Z$ bosons that enables one to calibrate the detector, both the tracking system and the calorimeters. One has to introduce proper corrections in the detector response in order to reconstruct the correct $Z$ mass. A similar strategy is used at proton (quark-gluon) colliders such as the LHC. For example, owing to $Z$-boson energy-scale calibration, the systematic error of 0.06\gevcc in the measurement of the Higgs boson mass was achieved with the ATLAS detector in the $ZZ^*$ decay mode based on 25 \invpb of integrated luminosity~\cite{LHC}.

   The cross section of the process $\ge \to eZ$ is rather large, only about a factor of 3 smaller than that for $\ge \to \ge$, see graphs for differential cross sections for these processes in ref.~\cite{Denner}.  The spectrum of colliding electrons and photons at photon colliders is very broad, so many $Z$ bosons will be produced with a low longitudinal momentum, which enables calibration of the whole detector at $P_\perp \sim 0.5~M_Z$. To ensure the linearity of detector response up to the maximum energies, one can use $Z$ bosons emitted at large angles. The cross section for such events is smaller than the total by a factor of $\ln(E_0/m_e) \approx 10$. Decays of $Z$ to leptons and jets allow all detector components to be calibrated.

\section{Conclusion}
    At the photon collider, the edge energy of the photon spectrum and the electron beam energy $E_0$ are not exactly related due to nonlinear effects in
    Compton scattering. For this reason, one cannot use the process $\ge \to \ge$ to measure the beam energy and calibrate the detector.  In addition, in \ge collisions, electrons are non-monochromatic due to large beamstrahlung (larger than in \epem due to the smaller horizontal beam size used at photon colliders). Due to these facts, precise knowledge of the initial beam is not necessary. In fact, only the absolute energy scale of the detector is needed. Such energy calibration can be done using the process $\ge \to eZ$, whose cross section is sufficiently large.

    A beam-energy spectrometer upstream of the interaction point (foreseen for \epem collisions) will be useful for fast determination of the initial beam energy, which would be particularly useful in experiments that involve beam-energy scanning. The downstream spectrometer is not possible at photon colliders due to highly disrupted beams.

\section*{Acknowledgments}

The work was supported by the Ministry of Education and Science of the Russian Federation.

\end{document}